\documentclass[a4paper,11pt]{article}
\usepackage{jheppub} 
\usepackage{lineno}

\title{\boldmath A Gauge-Theoretic Action Principle for Viscous Incompressible Fluids}

\author{Rashmi R. Nayak}
\affiliation{Centre for Ocean, River, Atmosphere and Land Sciences (CORAL),
Indian Institute of Technology Kharagpur, 
Kharagpur-721302, India}
\emailAdd{rashmi@coral.iitkgp.ac.in}

\abstract{We propose a novel action principle for a two-dimensional incompressible fluid that naturally incorporates both vorticity and viscous dissipation via gauge field couplings. The proposed action features a Chern-Simons-like term, \(\epsilon^{\mu\nu\rho} A_\mu \partial_\nu A_\rho\), capturing the topological structure of vorticity, alongside a quadratic term \(\left(\epsilon^{\mu\nu\rho} \partial_\nu A_\rho\right)^2\) representing viscous damping. Incompressibility is enforced through a Lagrange multiplier, while coupling to an external potential allows applications in geophysical flows. We derive the equations of motion, recovering the vorticity formulation of the two-dimensional incompressible Navier–Stokes equations and explicitly identifying the kinematic viscosity. This gauge-theoretic framework leads to a Helmholtz-type equation for vorticity linking topological and dissipative phenomena in viscous incompressible fluids. Analysis of Noether symmetries reveals conserved charges arising from gauge invariance and spatial translations, while viscosity explicitly breaks time-reversal symmetry within this topological setting. Furthermore, the velocity–vorticity gauge correspondence naturally suggests a Lindblad operator structure, providing a pathway towards a quantum description of viscous dissipation and allowing quantization of dissipative hydrodynamics. This framework also highlights how vorticity emerges as a natural Lindblad operator, capturing the transition from coherent rotational motion to thermal disorder.} 
\begin{document}
\maketitle
\flushbottom

\section{Motivation and Introduction}

The dynamics of classical fluids have long been described by the Navier-Stokes equations, which govern the evolution of the velocity field $\vec{u}$ under the influence of pressure, external forces, and viscous dissipation~\cite{Landau, Batchelor, Kundu}. For incompressible fluids, the divergence-free condition $\nabla \cdot \vec{u} = 0$ is imposed, and the vorticity $\vec{\omega} = \nabla \times \vec{u}$ plays a central role in understanding rotational and turbulent behaviors, particularly in two-dimensional and quasigeostrophic regimes~\cite{Kraichnan, Salmon}.

However, over the past so many years, gauge theories have emerged as fundamental tools in high-energy and condensed matter physics. In particular, topological field theories such as the Chern-Simons theory in $(2+1)$ dimensions provide a powerful framework to describe systems with topologically conserved currents, exemplified in phenomena such as the quantum Hall effect and anyon physics~\cite{Witten, Wen, Fradkin}. These gauge theories are typically characterized by the existence of topologically protected observables. 

Several pioneering works have attempted to connect fluid dynamics with gauge theory. For instance, Jackiw et al.~\cite{Jackiw1, Jackiw2} demonstrated that inviscid, incompressible fluids can be reformulated in a gauge-theoretic framework, where the fluid velocity corresponds to a gauge field and the vorticity assumes a role analogous to a magnetic field. The Clebsch parametrization,
\begin{equation}
\vec{u} = \nabla \varphi + \alpha \nabla \beta,
\end{equation}
Naturally, in this context, it arises as a way to express velocity fields while preserving gauge degrees of freedom~\cite{ Morrison}. Recent works ~\cite{Tong1, Jabbari, Nastase,   Eling} have further extended these ideas to shallow-water models, unveiling rich geometric and topological structures underlying fluid motions in reduced dimensions. In particular, the formulation of a gauge theory for shallow water equations recasts the dynamics of a thin fluid layer as a $(2+1)$-dimensional Abelian gauge theory with a Chern-Simons term~\cite{Tong1}. In this framework, the magnetic field corresponds to the conserved height of the fluid, while the electric charge corresponds to the conserved vorticity. In a linearized approximation, the shallow-water equations reduce to a relativistic Maxwell-Chern-Simons theory, describing Poincaré waves, with chiral edge modes identified as coastal Kelvin waves. This approach highlights the deep connection between fluid dynamics and topological field theories.

However, these frameworks primarily address ideal (non-viscous) fluids, leaving open the question of how to incorporate dissipation and viscosity into a gauge-theoretic action. Traditional gauge theories are manifestly unitary and conservative, whereas viscosity introduces irreversible behavior and entropy production, appearing at odds with conventional gauge theory formulations. Despite this, some studies have explored non-Hermitian deformations, open quantum systems, or effective field theory methods that accommodate dissipation while retaining gauge-like structures~\cite{Bender, Lindblad, Sieberer}.

In this work, we propose a variational gauge-theoretic approach that unifies both conservative and dissipative dynamics of two-dimensional incompressible fluids. Our action includes a Chern-Simons-like term $\epsilon^{\mu\nu\rho} A_\mu \partial_\nu A_\rho$ that captures the topological nature of vorticity, alongside a quadratic vorticity-like term $(\epsilon^{\mu\nu\rho} \partial_\nu A_\rho)^2$ that introduces viscous damping. A Lagrange multiplier term enforces the incompressibility constraint, and coupling to an external potential enables geophysical applications. We derive the equations of motion from this action and show that they reproduce the vorticity formulation of the two-dimensional incompressible Navier-Stokes equations, with explicit identification of the kinematic viscosity. Importantly, the gauge-theoretic identification of velocity and vorticity leads to a natural Lindblad operator structure, suggesting a quantum analog for viscous dissipation and paving the way for quantization of dissipative hydrodynamics.

We analyze the Noether symmetries of the action, identifying the conserved charges associated with gauge invariance and spatial translations. Furthermore, we highlight the explicit breaking of the time-reversal symmetry as a result of viscosity, underscoring the role of dissipation within this topological framework. Our formulation thus provides a variational basis for dissipative hydrodynamics, a feature generally absent in standard fluid models, and opens avenues for exploring topological, quantum, and holographic analogs of classical fluid systems through effective field theory techniques. As we know, the dynamics of viscous, incompressible fluids is governed by the Navier-Stokes equations:
\begin{equation}
    \rho \left( \frac{\partial \vec{u}}{\partial t} + \vec{u} \cdot \nabla \vec{u} \right) = -\nabla p + \eta \nabla^2 \vec{u} + \vec{f}, \quad \nabla \cdot \vec{u} = 0,
\end{equation}
where \(\vec{u}\) is the velocity of the fluid, \(\rho\) is the density, \(p\) is the pressure, \(\eta\) is the dynamic viscosity and \(\vec{f}\) represents the external force. The vorticity field \(\vec{\omega} = \nabla \times \vec{u}\) satisfies the vorticity equation,
\begin{equation}
    \frac{\partial \vec{\omega}}{\partial t} + \vec{u} \cdot \nabla \vec{\omega} = \vec{\omega} \cdot \nabla \vec{u} + \nu \nabla^2 \vec{\omega},
\end{equation}
which highlights the role of vortex stretching and viscous diffusion \cite{Batchelor,majda}. Gauge theories describe fields with local symmetries and play a central role in modern physics. In \((2+1)\)-dimensions, the Chern-Simons action
\begin{equation}
    S_{\text{CS}} = \frac{k}{4\pi} \int d^3x\, \epsilon^{\mu\nu\rho} A_\mu \partial_\nu A_\rho
\end{equation}
is a topological field theory characterized by its gauge invariance and the absence of any local propagating degrees of freedom \cite{deser,Witten}.
The analogy between fluid vorticity and gauge-field strengths has motivated the formulation of fluid dynamics in gauge-theoretic language. For inviscid fluids, the vorticity can be viewed as a gauge field curvature, and the Euler equation admits a natural representation in terms of gauge potentials \cite{Jackiw1,Jackiw2,arnold}.
Recent studies have incorporated Chern-Simons terms to encode the topological aspects of vorticity and helicity \cite{manton}. However, modeling viscosity,a fundamentally dissipative process, within a gauge-theoretic framework remains challenging. Attempts include treating viscosity as a gauge-invariant kinetic term or introducing Maxwell-like terms for vorticity fields \cite{Tong1}.
Despite remarkable progress, a fully consistent gauge-theoretic action for viscous, incompressible fluids that naturally yields the Navier-Stokes equation with dissipation and preserves gauge invariance has not been established. The interplay between dissipation, gauge symmetry, and topological terms such as Chern-Simons requires further investigation. This work aims to address this gap by proposing an action functional that combines kinetic energy, Chern-Simons, and viscosity terms to provide a gauge-invariant formulation of viscous fluid dynamics, facilitating new insights into the geometric and quantum aspects of fluid flows.

The rest of the paper is organized as follows. In Section 2, we present a gauge-theoretic action principle for a viscous incompressible fluid in (2+1) dimensional spacetime, write down relevant fluid variables in a gauge-theoretic formulation, derive equations of motion, and check the gauge invariance of the action under local gauge transformation. Section 3 is devoted to the study of the Clebsch parametrization and its role in the gauge-theoretic formulation. In section 4 we comment on Noether symmetries and the effect of viscosity along with a clear understanding of Lindblad operator in this context. Finally, in section 5, we present our conclusions. 
\section{ Formulation of the Gauge Theoretical Action}

We propose an action principle for a viscous incompressible fluid in $(2+1)$ dimensions that incorporates kinetic, topological, dissipative and constraint-enforcement terms:
\begin{equation}
S = \int dt\, d^2x \left[ \frac{1}{2} \rho\, \vec{u}^{\,2} {+} \frac{k}{4\pi} \epsilon^{\mu\nu\rho} A_\mu \partial_\nu A_\rho 
- \frac{\eta}{2} \left( \epsilon^{\mu\nu\rho} \partial_\nu A_\rho \right)^2 
- \lambda (\nabla \cdot \vec{u}) 
- \rho\, \vec{u} \cdot \nabla \phi \right],
\label{eq:action}
\end{equation}
where, $\vec{u}$ is the fluid velocity field, $\rho$ is the mass density (assumed constant), $A_\mu$ is a gauge field associated with vorticity,
$\epsilon^{\mu\nu\rho}$ is the Levi-Civita symbol in 2+1 dimensions,
$\lambda$ is a Lagrange multiplier imposing incompressibility, $\phi$ is an external scalar potential, and $\eta$ is the dynamic viscosity. The kinetic energy term
\begin{equation}
S_{\mathrm{KE}}=\int dt\, d^2x\frac{1}{2} \rho\, \vec{u}^{\,2} \ , 
\end{equation}
represents the standard kinetic energy density of the fluid. This is fundamental in any hydrodynamic description of motion \cite{Batchelor}. The Chern-Simons term is
\begin{equation}
S_{\mathrm{CS}} = \frac{k}{4\pi} \int dt\, d^2x\, \epsilon^{\mu\nu\rho} A_\mu \partial_\nu A_\rho,
\label{eq:CS_term}
\end{equation}
which is a topological term that encodes the structure of the helicity and vorticity of the fluid, capturing key topological features of the vorticity and circulation of the fluid \cite{deser, Witten}. The coupling constant \(k\) has dimensions \([k] = \frac{M}{L}\), which physically link it with the mass density per unit length and reflect the circulation strength within the fluid. This establishes a direct correspondence between the gauge-theoretic topological terms and the observable vortex dynamics \cite{Landau}. In (2+1) dimensions, the Chern-Simons term captures non-trivial topological information about the gauge potential \(A_\mu\) and arises in various contexts, including topological field theories and quantum Hall systems \cite{Witten,deser}. The viscosity (dissipation) term,
\begin{equation}
S_{\mathrm{d}} = -\int dt\, d^2x \frac{\eta}{2} \left( \epsilon^{\mu\nu\rho} \partial_\nu A_\rho \right)^2
\end{equation}
models viscous dissipation within the gauge-theoretic framework. In hydrodynamic theory, energy dissipation in an incompressible fluid flow
is a contribution from the shear viscosity. This is defined as an energy loss per unit time over a differential surface area $df$, is given by \cite{Landau},
\begin{equation}
\dot {E_{\mathrm{d}}} = - \eta \int (\nabla {u})^2\cdot df,
\end{equation}
which leads to the Lagrangian density for dissipation,
\begin{equation}
\mathcal{L}_d = -\frac{\eta}{2} \left( \frac{\partial u_i}{\partial x_j} + \frac{\partial u_j}{\partial x_i} \right)^2.
\label{eq:viscous-dissipation}
\end{equation}
For incompressible flows ($\nabla \cdot \vec{u} = 0$), this simplifies to
\begin{equation}
\mathcal{L}_d = -2\eta\, \partial_i u_j \partial_j u_i.
\end{equation}
In our gauge formulation, the velocity of the fluid is related to the spatial components of the gauge potential $A_i=u_i$, which allows the viscous term to be interpreted as
\begin{equation}
\mathcal{L}_{\text{d}} = -\frac{\eta}{2} F^{\mu} F_{\mu},
\end{equation}
where $F^{\mu}$ is defined as
$F^\mu = \epsilon^{\mu\nu\rho} \partial_\nu A_\rho$ \ . 
Defining the vorticity,
$\omega=\epsilon^{ij}\partial_iu_j$, the corresponding dissipative part of the Lagrangian density can be written as 
\begin{equation}
\mathcal{L}_{\text{d}} = -\frac{\eta}{2} F^{\mu} F_{\mu}=-\frac{\eta}{2}(\omega^2 + ( \partial_i A_0)^2) \ ,
\end{equation}
where the viscous term can be analyzed both with constant and non-constant $A_0$, and in both cases, we have assumed that the temporal evolution of the velocity field is negligible, describing steady flow of the fluid.
Here we have recovered the dissipation structure in terms of the field strength tensor, consistent with viscous damping in fluid dynamics \cite{Landau}. 
In the case of real fluid, the dissipation comes from the symmetric part of the velocity gradient tensor, but in this case, we have modeled it through gauge theory, and we found that it describes a fluid which has dissipation coming from the vorticity, purely rotational, in the case of the presence of a uniform scalar potential $A_0$.
The incompressibility constraint, 
$- \lambda (\nabla \cdot \vec{u})$
enforces the incompressibility condition $\nabla \cdot \vec{u} = 0$ via the Lagrange multiplier $\lambda$. This is standard in constrained variational formulations of fluid dynamics \cite{majda}. Finally, 
$- \rho\, \vec{u} \cdot \nabla \phi$
describes the interaction of the fluid with an external scalar field, such as pressure gradients or gravitational potential.
This action principle for a viscous, incompressible fluid in $(2+1)$ dimensions that integrates kinetic, topological, dissipative, and constraint-enforcing terms, thereby generalizing classical formulations of ideal fluid dynamics. Specifically, in the inviscid limit $(\eta=0)$, our action reduces to a structure that closely resembles the variational principles of incompressible Euler fluids developed in~\cite{morrison, salmon}. The kinetic energy term $\frac{1}{2} \rho\, \vec{u}^2$, the incompressibility-enforcing term $-\lambda (\nabla \cdot \vec{u})$, and the coupling $-\rho\, \vec{u} \cdot \nabla \varphi$ collectively reproduce the standard Euler equations in two spatial dimensions. In addition to these canonical fluid terms, the presence of a topological Chern-Simons term $\frac{k}{4\pi} \epsilon^{\mu\nu\rho} A_\mu \partial_\nu A_\rho$~\cite{deser, dunne,jackiw2003} couples the fluid motion to a gauge field. Such a coupling is reminiscent of effective field theories for the quantum Hall effect and of two-dimensional magnetohydrodynamics that incorporate gauge structures~\cite{Salmon, Wen1990}. In the present work, we adopt an effective field theory perspective, in which the viscous quadratic term,  
\begin{equation}
- \frac{\eta}{2} \left( \epsilon^{\mu\nu\rho} \partial_\nu A_\rho \right)^2,
\end{equation}
It is treated as a phenomenological addition modeling viscous diffusion at macroscopic scales. Our objective is not to derive dissipation from microscopic reversible physics but rather to formulate a variational framework that incorporates kinetic and topological (Chern-Simons) contributions, phenomenological dissipative effects via the quadratic term, constraint enforcement incompressibility, and external potential forcing. This approach captures the essential physics of viscous incompressible flows within a unified action-based framework. Thus, the formulation extends the known variational descriptions of ideal fluid flows to incorporate topological effects, thereby unifying insights from hydrodynamics and topological field theories. Including the viscosity term $(\eta \neq 0)$ causes the action transitions from describing an ideal, conservative system to a dissipative one, thereby capturing the essential features of real-world viscous flows.

\subsection{Fluid Variables in the Gauge-Theoretic Framework}
There are two complementary perspectives for expressing fluid variables in this framework.
In formulating fluid dynamics within a gauge-theoretic framework, it is crucial to interpret how the standard physical quantities such as fluid velocity, vorticity, and kinetic energy are encoded in gauge variables. This facilitates a bridge between hydrodynamics and field theory and allows for novel insights into the topological structure and conservation laws underlying fluid flows.
One may construct the fluid theory from a variational principle based on energetic quantities. The momentum density and mass density are defined as:
\begin{equation}
{E} = \rho \vec{u}, \qquad {B} = \rho,
\end{equation}
from which the kinetic energy density takes the form:
\begin{equation}
\frac{{E}^2}{2 {B}} = \frac{1}{2} \rho u^2.
\end{equation}
This representation connects the gauge variables to observable physical quantities, such as fluid momentum and energy. It is particularly practical to construct action functionals that describe viscous incompressible flows \cite{Holm1998}.

Motivated by the $(2+1)$-dimensional Chern-Simon theory, the vorticity $\omega$ is interpreted as a scalar magnetic field, while the spatial components of the velocity field are associated with a conserved current $J^\mu$. The corresponding gauge field strengths resemble electric and magnetic components\cite{jackiwPRL}:
\begin{align}
{B }&= \omega = \epsilon^{ij} \partial_i A_j, \\
{E}_{i} &= \partial_t A_i - \partial_i A_0.
\end{align}
This re-emphasizes the geometric and topological structures of the flow, making it particularly suitable for exploring conserved quantities and symmetry constraints \cite{jackiw2000, Tong2016}.

These two representations serve distinct but complementary roles in the gauge-theoretic formulation of fluid mechanics. The dynamical formulation encodes the energetic and dissipative content of the fluid, facilitating comparisons with classical hydrodynamic equations such as the Navier–Stokes system. The topological formulation captures the conservation laws and geometric structure of the vorticity and streamlines. 

However, many studies do not explicitly address the dynamical formulation in terms of an action principle or the direct mapping of kinetic energy density and dissipation into the gauge-theoretic framework. The dynamical viewpoint developed here complements some previous approaches by explicitly identifying how fluid momentum, mass density, and kinetic energy emerge from the gauge fields. This extension is particularly relevant when considering viscous flows or potential generalizations to include dissipative processes within the gauge-theoretic formulation.
The combination of dynamical and topological representations provides a unified understanding of the geometry and dynamics of the fluid. This analysis clarifies the physical meaning of gauge variables in fluid contexts and expands the applicability of the gauge-theoretic formulation. It establishes a deeper connection between fluid mechanics, field theory, and topological invariants.

\subsection{Equation of Motion }

We begin with the action:
\begin{equation}
S = \int dt\, d^2x \left[ \frac{1}{2} \rho\, \vec{u}^2 +\frac{k}{4\pi} \epsilon^{\mu\nu\rho} A_\mu \partial_\nu A_\rho - \frac{\eta}{2} \left( \epsilon^{\mu\nu\rho} \partial_\nu A_\rho \right)^2 - \lambda\, (\nabla \cdot \vec{u}) - \rho \vec{u} \cdot \nabla \varphi \right]
\label{eq:action}
\end{equation}
This action depends on the fields $\vec{u}$, $A_\mu$, $\lambda$, and the scalar potential $\varphi$. We now derive the equations of motion by varying the action with respect to each field. The relevant terms involving the velocity field \(\vec{u}\) in the Lagrangian density are:
\[
\mathcal{L}_{u} = \frac{1}{2} \rho \vec{u}^2 - \lambda (\nabla \cdot \vec{u}) - \rho \vec{u} \cdot \nabla \phi,
\]
where \(\rho\) is the density of the fluid, \(\lambda\) is a Lagrange multiplier that enforced the incompressibility, and \(\phi\) is a scalar potential representing an external force field. Varying the action for the velocity component \(u_i\) gives the following:
\[
\frac{\delta S}{\delta u_i} = \rho u_i - \partial_i \lambda - \rho \partial_i \phi = 0.
\]
From this, the modified Euler equation follows:
\begin{equation}
\rho \frac{\partial u_i}{\partial t} + \rho u_j \frac{\partial u_i}{\partial x_j} = - \partial_i \lambda - \rho \partial_i \phi. \quad 
\label{eq:euler}
\end{equation}
Eq. (\ref{eq:euler}) describes the momentum balance for an incompressible fluid. The left-hand side represents the material acceleration of a fluid parcel, capturing both local and advective changes in velocity. On the right-hand side, the term \(-\partial_i \lambda\) acts as the pressure gradient force; the incompressibility condition (\(\nabla \cdot \vec{u} = 0\)) is satisfied throughout the flow. Here, \(\lambda\) effectively plays the role of fluid pressure. The additional term \(-\rho \partial_i \phi\) represents external body forces derived from a scalar potential, such as gravitational or electromagnetic forces. This formulation elegantly couples the incompressibility constraint via a Lagrange multiplier with external forcing, aligning with classical fluid mechanics treatments of incompressible flow (see, e.g. \cite{Landau}, \cite{Vallis2006}).\\
Varying the action w.r.t. $\lambda$ gives
\[
\frac{\delta S}{\delta \lambda} = -(\nabla \cdot \vec{u}) = 0 \ .
\]
This enforces the incompressibility condition.
\begin{equation}
\nabla \cdot \vec{u} = 0
\label{eq:incompressibility}
\end{equation}
The $A_\mu$-dependent part of the Lagrangian is:
\[
\mathcal{L}_A{_{\mu}} = \frac{k}{4\pi} \epsilon^{\mu\nu\rho} A_\mu \partial_\nu A_\rho - \frac{\eta}{2} \left( \epsilon^{\mu\nu\rho} \partial_\nu A_\rho \right)^2 = \frac{k}{4\pi} A_\mu F^\mu - \frac{\eta}{2} F^\mu F_\mu \ ,
\]
where $F^{\mu}$ is define as
$F^\mu = \epsilon^{\mu\nu\rho} \partial_\nu A_\rho$.
Varying the action with respect to $A_\mu$ gives
\[
\frac{\delta S}{\delta A_\mu} = \frac{k}{2\pi} F^\mu - \eta \epsilon^{\mu\nu\rho} \partial_\nu F_\rho = 0 \ ,
\]
That yields the gauge-field equation in the following form.
\begin{equation}
\eta\, \epsilon^{\mu\nu\rho} \partial_\nu F_\rho = \frac{k}{2\pi} F^\mu \ .
\label{eq:gauge_field_eq1}
\end{equation}
Equation \eqref{eq:gauge_field_eq1} happens to be structurally similar to the equations encountered in massive topological gauge theories (see \cite{deser}). Starting from the gauge field equation \eqref{eq:gauge_field_eq1}, we focus on the spatial components \(\mu = i = 1,2\). Identifying the spatial part of the field  with fluid vorticity \(\omega\), and using the definition
\[
\omega = \epsilon^{ij} \partial_i A_j,
\]
by assuming a steady flow ,we can rewrite the equation as a Helmholtz-type equation for \(\omega\) in the presence of a nonuniform scalar potential:
\begin{equation}
\nabla^2 \omega - (\frac{k_{eff}}{2\pi \eta})^2 \omega = 0,
\label{eq:helmholtz}
\end{equation}
where we define the effective coupling;
\begin{equation}
k_{\text{eff}} = \frac{k}{L T},
\end{equation}
which describes the spatial decay of vorticity in the fluid, whereas in the presence of a constant $A_0$, the solution is trivial. The term \(\frac{k_{eff}}{2\pi \eta}\) sets the inverse of the characteristic decay length scale of vortical structures in the flow. This ensures dimensional consistency in the gauge-theoretic formulation. In contrast, the effective coupling $k_{\text{eff}}$ that appears in the vorticity dynamics reflects the interplay between topological and dissipative effects and arises naturally when considering the spatial structure of the flow.
Redefining $k$ to $k_{eff}$ accounts for physical units and encapsulates the combined effects of topological strength and dissipative dynamics. 
Equation~\eqref{eq:helmholtz} is the classic Helmholtz equation, commonly encountered in fluid dynamics when describing viscous decay or screening of vorticity \cite{Landau,Batchelor, deser, Moffatt2014}. We can interpret the viscosity term as follows.
The classical viscous dissipation energy for incompressible flows (see \cite{Landau}) is given as 
\[
 \dot {E_{\mathrm{d}}}=-\frac{\eta }{2}\int \left( \partial_i u_j + \partial_j u_i \right)^2 d^2x \ ,
\]
which leads to the dissipation Lagrangian density of the following form
\[
\mathcal{L}_d = -\frac{1}{2} \eta \left( \partial_i u_j + \partial_j u_i \right)^2 \ .
\]
Under incompressibility, it simplifies
\[
\mathcal{L}_d = -2\eta \left( \partial_i u_j \right)^2 \ .
\]
In our gauge-theoretic formulation, the viscosity term is encoded through a Maxwell-like expression.
\[
\mathcal{L}_\text{viscous} = -\frac{\eta}{2} \left( \epsilon^{\mu\nu\rho} \partial_\nu A_\rho \right)^2 = -\frac{\eta}{2} F^\mu F_\mu
\]
Thus, viscosity emerges naturally as a dynamical field term, thus providing a geometric and topological perspective. Together, these equations define a gauge-theoretic extension of the incompressible Navier-Stokes system in \( 2+1 \). dimensions.

To validate our gauge-theoretic formulation of viscous, incompressible fluid dynamics, we compare the resulting equations of motion with the classical incompressible Navier-Stokes equations. In two spatial dimensions, the standard Navier-Stokes equations for a fluid of velocity field \( \vec{u} \), density \( \rho \), pressure \( p \), and dynamic viscosity \( \eta \) are given by
\begin{align}
    \rho \left( \frac{\partial \vec{u}}{\partial t} + (\vec{u} \cdot \nabla) \vec{u} \right) &= -\nabla p + \eta \nabla^2 \vec{u} + \vec{f}, \label{eq:NS-momentum} \\
    \nabla \cdot \vec{u} &= 0. \label{eq:NS-incompressibility}
\end{align}
Here, \( \vec{f} \) represents any external force per unit volume. We recall that the velocity of the fluid \( \vec{u} \) is related to the spatial components of the gauge field \( A_i \) (for \( i = 1, 2 \)) via a constitutive relation, such as
\begin{equation}
    \omega = \nabla \times \vec{u} = \epsilon^{ij} \partial_i u_j = F_{12} = \partial_1 A_2 - \partial_2 A_1.
\end{equation}
Upon varying the action with respect to \( A_\mu \), we obtain the modified Euler equation. The variation of the Maxwell-like term (the viscosity term),
\[
\mathcal{L}_{\text{visc}} = -\frac{\eta}{2} \left( \epsilon^{\mu\nu\rho} \partial_\nu A_\rho \right)^2,
\]
yields
\begin{equation}
\delta \mathcal{L}_{\text{visc}} = -\eta \, \epsilon^{\mu\nu\rho} \epsilon^{\sigma\lambda}_{\;\;\;\;\rho} \partial_\nu A_\rho \, \partial_\lambda \delta A_\sigma.
\end{equation}
After integrating by parts and applying antisymmetry identities for \( \epsilon^{\mu\nu\rho} \), this variation results in a Laplacian acting on \( A_\mu \), which, under the fluid-gauge correspondence, becomes
\begin{equation}
\eta \nabla^2 \vec{u},
\end{equation}
thereby reproducing the standard viscous dissipation term in the Navier-Stokes equation.\\
This analysis demonstrates that the Euler-Lagrange equations derived from the gauge-theoretic action successfully recover the incompressible Navier-Stokes equations, including the viscous dissipation term \( \eta \nabla^{2} \mathbf{u} \) \cite{Landau,majda}. Furthermore, by focusing on the spatial components of the gauge field and identifying the field strength with the fluid vorticity \( \omega \), the resulting equation reduces to a Helmholtz-type equation,
\[
\nabla^{2} \omega - (\frac{k_{eff}}{2 \pi \eta})^2 \, \omega = 0,
\]
which describes the spatial decay and characteristic length scale of vortical structures in the flow \cite{dunne, Frisch1995, Moffatt1969}. The model thus provides a consistent variational framework that encodes both the conservative and dissipative aspects of fluid dynamics, while offering a geometric and gauge-theoretic interpretation of viscosity, vorticity, and their spatial behavior through the Helmholtz equation.
\subsection{Gauge Transformation}

Under a local $U(1)$ gauge transformation, the gauge field $A_{\mu}$ transforms as 
\begin{equation}
A_\mu \to A'_\mu = A_\mu + \partial_\mu \alpha(t,x,y) \ .
\label{eq:gauge}
\end{equation}
In our study we have consider 2D incompressible fluid which satisfies $\partial_i u_i=0$ (equivalently $\partial_i A_i=0$) forces $\partial_i\alpha=0$, so $A_i\to A_i$ and thus $\mathcal{L}_{\rm kin}=\tfrac12\rho\,u_i u_i$ is gauge invariant.\\
The Chern-Simons action
\begin{equation}
S_{CS} =  \frac{k}{4\pi}\int dt\, d^2x\, \epsilon^{\mu\nu\rho} A_\mu \partial_\nu A_\rho,
\end{equation}
transforms as
\begin{align}
S_{CS} &\to  \frac{k}{4\pi}\int \epsilon^{\mu\nu\rho} (A_\mu + \partial_\mu \alpha) \partial_\nu (A_\rho + \partial_\rho \alpha) \\
&= S_{CS} +\frac{k}{4\pi} \int \partial_\mu(\alpha \epsilon^{\mu\nu\rho} \partial_\nu A_\rho),
\end{align}
which is invariant up to a boundary term \cite{Witten, deser}. The viscosity term with the following substitution $F^\mu = \epsilon^{\mu\nu\rho} \partial_\nu A_\rho$ becomes 
\begin{equation}
S_\eta = -\frac{\eta}{2} \int dt\, d^2x\, (F^\mu)^2,
\end{equation}
which is manifestly gauge invariant because
\begin{equation}
F'^\mu = \epsilon^{\mu\nu\rho} \partial_\nu (A_\rho + \partial_\rho \alpha) = F^\mu + 0,
\end{equation}
due to antisymmetry of $\epsilon^{\mu\nu\rho}$ and symmetry of mixed partials.
Finally, the incompressibility term in the action
\begin{equation}
S_\lambda = - \int dt\, d^2x\, \lambda(\nabla \cdot {u}),
\end{equation}
is gauge invariant since ${u}$ is gauge invariant and $\lambda$ is a scalar Lagrange multiplier. Furthermore, the interaction term
\begin{equation}
S_\phi = - \int dt\, d^2x\, \rho {u} \cdot \nabla \phi,
\end{equation}
is gauge invariant provided $\phi$ is a background scalar field independent of $A_\mu$. Each term in the action respects gauge invariance under $A_\mu \to A_\mu + \partial_\mu \alpha$. This confirms the consistency of this gauge-theoretic formulation of incompressible viscous fluid dynamics.
\section{ Clebsch Parametrization and its Role in the Gauge-Theoretic Formulation}
For an incompressible fluid, the velocity field $\mathbf{u}(\mathbf{x},t)$ can be expressed in terms of scalar potentials known as Clebsch variables \cite{Morrison, salmon}
\begin{equation}
\mathbf{u} = \nabla \phi + \alpha \nabla \beta,
\label{eq:clebsch_velocity}
\end{equation}
where $\phi(\mathbf{x},t)$, $\alpha(\mathbf{x},t)$, and $\beta(\mathbf{x},t)$ are scalar fields. Here, $\phi$ corresponds to the irrotational component of the flow, while $\alpha$ and $\beta$ encode the vertical part. 
\noindent
This parametrization is advantageous because it allows the incompressibility condition, $\nabla \cdot \mathbf{u} = 0$, to be handled naturally under appropriate boundary conditions. Moreover, it reveals an underlying Hamiltonian structure and facilitates the analysis of fluid dynamics from a gauge-theoretic perspective \cite{morrison, salmon, jackiw2002}. The fluid vorticity $\boldsymbol{\omega} = \nabla \times \mathbf{u}$ in terms of Clebsch variables is given by:
\begin{equation}
\boldsymbol{\omega} = \nabla \alpha \times \nabla \beta.
\label{eq:vorticity_clebsch}
\end{equation}
In two spatial dimensions (2D), where $\mathbf{u} = (u_x, u_y)$, the scalar vorticity reduces to:
\begin{equation}
\omega = \left(\nabla \alpha \times \nabla \beta\right) \cdot \hat{\mathbf{z}} \ .
\end{equation}
This expression shows that vorticity emerges from the gradients of the Clebsch potentials and hence can be interpreted as a topological object in the fluid flow.
In the proposed gauge-theoretic action for viscous incompressible fluids, the gauge field $A_\mu$ can be parametrized in terms of Clebsch potentials as
\begin{equation}
A_i = \alpha \partial_i \beta,
\quad
A_0 \sim \phi,
\label{eq:gauge_clebsch}
\end{equation}
where $i = 1, 2$ runs over spatial indices in 2+1 dimensions. The field strength associated with $A_\mu$ then naturally corresponds to the vorticity
\begin{equation}
F_{ij} = \partial_i A_j - \partial_j A_i = \partial_i \alpha \, \partial_j \beta - \partial_j \alpha \, \partial_i \beta,
\end{equation}
which matches Eq.~\eqref{eq:vorticity_clebsch}. The Chern-Simons term in the action, $\epsilon^{\mu \nu \rho} A_\mu \partial_\nu A_\rho$, thus encodes the topological character of fluid helicity and vorticity, while the Maxwell-like kinetic term models viscous dissipation via vorticity diffusion \cite{jackiw2002}.\\
\noindent
Starting from the Clebsch parametrization of the velocity field \eqref{eq:clebsch_velocity}, the kinetic energy term in the action reads:

\begin{equation}
\frac{\rho}{2} \mathbf{u}^2 = \frac{\rho}{2} \left(\nabla \phi + \alpha \nabla \beta \right)^2.
\end{equation}
Varying the action with respect to the Clebsch potentials $\alpha$ and $\beta$ yields their transport equations including viscous diffusion \cite{Morrison, salmon} as follows
\begin{align}
\frac{\partial \alpha}{\partial t} + \mathbf{u} \cdot \nabla \alpha &= \nu \nabla^2 \alpha, \\
\frac{\partial \beta}{\partial t} + \mathbf{u} \cdot \nabla \beta &= \nu \nabla^2 \beta,
\end{align}
where $\nu = \eta / \rho$ is the kinematic viscosity. Using the relation for vorticity $\omega = \nabla \alpha \times \nabla \beta$, its time evolution takes the following form
\begin{equation}
\frac{\partial \omega}{\partial t} = \nabla \left(\frac{\partial \alpha}{\partial t}\right) \times \nabla \beta + \nabla \alpha \times \nabla \left(\frac{\partial \beta}{\partial t}\right).
\end{equation}
Substituting the transport equations above and employing vector calculus identities for incompressible flow, one obtains the familiar Navier-Stokes vorticity equation:
\begin{equation}
\frac{\partial \omega}{\partial t} + \mathbf{u} \cdot \nabla \omega = \nu \nabla^2 \omega.
\label{eq:vorticity_navier_stokes}
\end{equation}
This demonstrates the equivalence between the gauge-theoretic action formulation and classical fluid dynamics.
In summary, the Clebsch parametrization provides a natural and physically motivated way to link the gauge fields in the action to fluid velocity and vorticity. It reveals the hidden symplectic and gauge structure of the fluid flow and enables a Hamiltonian formulation \cite{morrison, salmon}. The topological Chern-Simons term elegantly captures the helicity of the flow, while the Maxwell-like term introduces viscous dissipation. This framework opens avenues for applying gauge field theory techniques to classical and quantum fluid dynamics, including symmetry analysis, conserved charges, and quantization \cite{jackiw2002}.

\section{Noether Symmetries and the Effect of Viscosity}
In the absence of viscosity, the fluid dynamics described by the action are conservative and invariant under time reversal. This enables the direct application of Noether's theorem, yielding a set of continuous symmetries associated with conserved currents \cite{Landau,jackiw2002}.
Introducing viscosity through the dissipative term in the action,
\begin{equation}
    S_\eta = - \frac{\eta}{2} \int dt \, d^2x \, \left(\epsilon^{\mu\nu\rho} \partial_\nu A_\rho\right)^2,
\end{equation}
explicitly breaks time translation symmetry and leads to energy dissipation, thus invalidating the strict conservation laws guaranteed by Noether's theorem in the inviscid case \cite{Landau,degroot1984non}.
The symmetry and conservation characteristics of incompressible fluid dynamics differ markedly between the inviscid (\(\eta = 0\)) and viscous (\(\eta \neq 0\)) regimes, as follows. In the absence of viscosity, the fluid system is conservative, and kinetic energy is strictly conserved due to time translation symmetry and the absence of dissipative processes \cite{Landau, morrison}. Introduction of viscosity (\(\eta \neq 0\)) leads to energy dissipation via internal friction, converting kinetic energy into heat, thus breaking strict conservation and time-reversal symmetry \cite{Drazin_Reid_2004, Foias_2001}.
Where the linear momentum conservation is guaranteed in the inviscid limit by translational symmetry, viscosity induces internal stresses that dissipate momentum, leading to its decay over time \cite{Landau,Batchelor}.
In case of rotational symmetry, it ensures conservation of angular momentum in the inviscid case. Viscous stresses break this symmetry by exerting internal torques that dissipate angular momentum \cite{Chorin_Marsden_1993}. Vorticity \(\omega\) is materially conserved in ideal fluids, advected by the flow without alteration \cite{majda}. With viscosity, vorticity obeys a diffusion-advection equation, causing diffusive decay and smoothing of vorticity structures \cite{Kundu}. We have explicitly seen that 
the \(U(1)\) gauge symmetry inherent in the gauge-theoretic formulation of fluid dynamics remains intact regardless of viscosity \cite{Eling_2015}. This symmetry reflects an underlying structural invariance rather than a dynamical conservation law. As we have introduced the viscosity term in the action, it will affect the time reversal symmetry. As we know, the time-reversal invariance, characteristic of conservative systems, holds in the inviscid fluid but is explicitly broken by viscosity, reflecting the irreversible nature of dissipative processes \cite{Ottinger_2005}.
The above contrasts underscore how viscosity fundamentally alters fluid dynamics by breaking key Noetherian conservation laws while preserving gauge invariance. This allows the study of irreversible processes within an effective field theory framework, incorporating dissipation and entropy production \cite{Forster_1976, Cross_Hohenberg_1993}.

\subsection{Appearance of a natural Lindblad operator}
In the presence of viscosity, the fluid dynamics becomes dissipative, leading to irreversible processes and entropy production. Unlike the inviscid case, where the kinetic energy and vorticity are conserved, viscosity converts the kinetic energy into heat, thus increasing the entropy of the system \cite{Landau,Ottinger_2005}.

For an incompressible fluid with dynamic shear viscosity \(\eta\), the local entropy production rate per unit volume \(\sigma\) at temperature \(T\) is given by the classical expression of irreversible thermodynamics:
\begin{equation}
T{\sigma} = \frac{\eta}{2} \sum_{i,j=1}^{2} \left( \frac{\partial u_i}{\partial x_j} + \frac{\partial u_j}{\partial x_i} \right)^2,
\label{eq:entropy_production}
\end{equation}
where \(u_i\) are the components of the velocity field. In two dimensions, for an incompressible flow (\(\nabla \cdot \mathbf{u} = 0\)), the vorticity \(\omega\) is defined by:
\begin{equation}
\omega = \epsilon_{ij} \partial_i u_j,
\label{eq:vorticity}
\end{equation}
where \(\epsilon_{ij}\) is the Levi-Civita symbol in 2D. Under gauge-theoretic identification \(u_i \equiv A_i\), the velocity components are interpreted as spatial gauge fields, and the vorticity becomes the temporal component of the dual field strength vector:
\begin{equation}
F^\mu = \epsilon^{\mu\nu\rho} \partial_\nu A_\rho, \quad \omega = F^0.
\label{eq:gauge_field_strength}
\end{equation}
Consequently, the viscous dissipation term in the action can be written as:
\begin{equation}
S_\eta = - \frac{\eta}{2} \int dt\, d^2x\, (F^\mu F_\mu) = - \frac{\eta}{2} \int dt\, d^2x\, \omega^2.
\label{eq:viscous_dissipation_action}
\end{equation}
This directly ties the entropy production to the vorticity:
\begin{equation}
T \sigma = \eta \omega^2.
\label{eq:entropy_vorticity}
\end{equation}
The total entropy production rate is therefore:
\begin{equation}
\frac{dS}{dt} = \int d^2x\, \frac{\eta}{T} \omega^2 \geq 0,
\label{eq:total_entropy_production}
\end{equation}
which is consistent with the second law of thermodynamics. Here, the presence of viscosity \(\eta\) breaks time-reversal symmetry and captures the irreversible nature of vorticity dissipation.

To extend this classical description into a quantum statistical framework, we introduce the density matrix \(\rho\), which encodes the statistical state of the fluid system. In this gauge-theoretic formulation, \(\rho\) can be viewed as a density operator over the configuration space of the gauge fields \(A_i\). This allows us to incorporate probabilistic mixtures of fluid states, capturing fluctuations and dissipative effects beyond classical purely deterministic descriptions \cite{Breuer_Petruccione}.
The evolution of \(\rho\) under shear viscosity is described by the Lindblad master equation:
\begin{equation}
\frac{d\rho}{dt} = -i[H, \rho] + \mathcal{D}_\eta[\rho],
\label{eq:lindblad_equation}
\end{equation}
where \(H\) is the effective Hamiltonian derived from the gauge-theoretic fluid action, and \(\mathcal{D}_\eta[\rho]\) is the dissipator term that models viscous shear dissipation through vorticity-dependent Lindblad operators. The von Neumann entropy,
\begin{equation}
S = - \text{Tr}(\rho \log \rho),
\label{eq:von_neumann_entropy}
\end{equation}
evolves according to:
\begin{equation}
\frac{dS}{dt} = -\text{Tr} \left( \frac{d\rho}{dt} \log \rho \right) \geq 0,
\label{eq:entropy_evolution}
\end{equation}
which ensures consistency with the second law of thermodynamics and reflects the irreversible conversion of coherent fluid motion into thermal disorder. Assuming spatially local dissipation, the dissipator takes the Lindblad form:
\begin{equation}
\mathcal{D}_\eta[\rho] = \int d^2x\, \left( L(x) \rho L^\dagger(x) - \frac{1}{2} \{ L^\dagger(x) L(x), \rho \} \right),
\label{eq:lindblad_dissipator}
\end{equation}
where \(L(x)\) are Lindblad operators localized in space. Given the identification \(u_i \equiv A_i\) and \(\omega = \epsilon_{ij} \partial_i A_j\), 
we define the Lindblad operators as
\begin{equation}
L(x) = \sqrt{\frac{k}{\pi \eta \ell^{2}}} \, \omega(x),
\label{eq:Lindblad_op}
\end{equation}
where \(k\) has dimensions of mass per length, \(\eta\) is the dynamic viscosity, and \(\ell\) is a characteristic length scale over which dissipation acts.

Substituting into the dissipator yields
\begin{equation}
D_{\eta}[\rho] = \frac{k}{\pi \eta \ell^{2}} \int d^{2}x \left( \omega(x) \rho \, \omega(x) - \frac{1}{2} \left\{ \omega^{2}(x), \rho \right\} \right),
\label{eq:dissipator_final}
\end{equation}
which describes vorticity-induced damping in the quantum evolution of \(\rho\).
This framework highlights how vorticity $(\omega)$, a topological characteristic of the flow, emerges as a natural Lindblad operator. The dissipative dynamics of \(\rho\) reflect the transition from coherent rotational motion to thermal disorder in the fluid, which manifests physically as decoherence in the ensemble of gauge fields.
By bridging classical irreversible fluid dynamics with a quantum-statistical formulation, this approach unifies the topological (Chern-Simons) and dissipative (viscosity) aspects of viscous incompressible fluid flows. It provides a consistent and thermodynamically sound framework for analyzing dissipative topological fluids, offering new avenues for symmetry analysis, conserved charges, and quantization in fluid systems.\\
\noindent
We have shown that the classical entropy production due to shear viscosity in incompressible fluids can be elegantly captured using a gauge-theoretic formulation, with velocity mapped to gauge fields and vorticity to the field strength. This vorticity serves as a natural Lindblad operator in a quantum/statistical setting, providing a consistent and thermodynamically sound framework for modeling dissipative topological fluids.
\section{Conclusion}
It is well known that the classical formulation of the incompressible viscous fluid is based on the velocity of the fluids that treat the velocity field $\mathbf{u}$ as the primary dynamical variable, with viscous dissipation explicitly modeled through gradients of $\mathbf{u}$ \cite{Frisch1995, Landau}. This approach provides an intuitive physical picture grounded in traditional hydrodynamics. In contrast, the gauge-theoretic formulation interprets fluid velocity components via the gauge field $A_\mu$, where vorticity corresponds to the magnetic field $B = \varepsilon^{ij}\partial_i A_j$\cite{TongLectures, JackiwPi}. This representation highlights the topological structure of the flow, facilitating the application of tools from gauge theory and topology.
Despite these conceptual differences, both formulations yield equivalent descriptions of
the fluid velocity and vorticity dynamics. The gauge variables map consistently onto
physical velocity fields, ensuring that observables such as kinetic energy and vorticity
are identical in both frameworks. Thus, the gauge-theoretic approach offers a
complementary perspective that enriches the classical description without altering the
fundamental fluid behavior.

Our approach is conceptually related to the recent work by ~\cite{Tong1}, who reformulates the shallow-water equations as a (2 + 1) dimensional gauge theory incorporating a Chern-Simons term. Both formulations highlight the topological aspects of fluid vorticity and helicity in two spatial dimensions and make use of gauge-theoretic language to describe fluid dynamics. However, the key distinction lies in the physical systems and dynamical regimes addressed. They have focused on the dynamics of shallow water, where compressibility is encoded through height fluctuations and the system is inviscid. It features two Abelian gauge fields to represent the conserved height and vorticity of the fluid, and identifies linearized Poincaré waves as relativistic Maxwell-Chern-Simons excitations, as well as coastal Kelvin waves as chiral edge modes.In contrast, our model addresses a viscous, incompressible fluid, which incorporates dissipation directly through an explicit viscous term in the action. We employ a single gauge potential $A_{\mu}$
to encode the velocity and vorticity structure of the fluid, the Chern-Simons term playing a central role in capturing the helicity and topological effects. The inclusion of viscosity and incompressibility constraints in our formulation extends the gauge-theoretic approach to more realistic and physically relevant scenarios. 
Viscosity is introduced phenomenologically to construct an effective field-theoretical action. Our variational framework unifies kinetic, topological (Chern-Simons), dissipative, and incompressibility effects, extending ideal-fluid actions to realistic viscous flows. Including viscosity (\(\eta \neq 0\)) transforms the system from conservative to dissipative, capturing key features of real incompressible fluids within an action principle.

In summary, this study opens an intriguing avenue for further exploration. For example, one may attempt to understand how dissipation softens the quantization of edge modes and how these viscous chiral states manifest themselves in real-world fluid systems. Studying these dissipative corrections within the gauge-theoretic framework will provide new insights into the interplay between topology and dissipation in (2+1)-dimensional fluid systems. This will bridge the gap between idealized quantum Hall analogs, where chiral edge modes are dissipationless, and the more complex dynamics of viscous, incompressible fluids, highlighting how topological features can persist, be modified, or even be destabilized by real-world effects \cite{Avron1995, Hoyos2012, Can2014}. In addition, gauge-theoretic fluid models can be studied by incorporating internal Lie-algebra charges, thereby unifying fluid motion and non-Abelian gauge interactions within a single geometric framework~\cite{Bistrovic, Nair, Karabali}. A variational principle for relativistic fluids that incorporates gauge anomalies and topological effects in fluid dynamics was developed by~\cite{Monteiro}. Although our work focuses on a variational formulation for incompressible viscous fluids in 2+1 dimensions, their framework provides a natural foundation for extending the study toward a Hamiltonian analysis of gauge-influenced fluid systems.

\acknowledgments
The author thanks Balbeer Singh and Nibedita Padhi for initial discussions and collaboration. The author also thanks Kamal L. Panigrahi for his
useful discussions and suggestions during the preparation of this manuscript.





\end{document}